\documentclass[prb,aps,floats,floatfix,twocolumn,epsf,showpacs,longbibliography]{revtex4-1}

\usepackage[USenglish]{babel}

\usepackage{amsmath,amssymb,amsfonts}

\usepackage{epsfig}
\usepackage{subfigure}

\usepackage{color}
\usepackage[dvipsnames]{xcolor}

\usepackage[colorlinks=true,linkcolor=blue,citecolor=red]{hyperref}

\newcommand{\eq}[2]
{
  \begin{equation}
    #1
    \label{#2}
  \end{equation}
}

\newcommand{\eqsplit}[2]
{
  \begin{equation}
    \begin{split}
      #1
    \end{split}
    \label{#2}
  \end{equation}
}

\newcommand{\equ}[1]
{Eq.~(\ref{#1})}

\newcommand{\figu}[1]
{Fig.\ref{#1}}

\newcommand{\secu}[1]
{Sec.~\ref{#1}}

\newcount\colveccount
\newcommand*\colvec[1]{
  \global\colveccount#1
  \begin{pmatrix}
    \colvecnext
  }
  \def\colvecnext#1{
    #1
    \global\advance\colveccount-1
    \ifnum\colveccount>0
    \\
    \expandafter\colvecnext
    \else
  \end{pmatrix}
  \fi
}

\newtoks\rowvectoks
\newcommand{\rowvec}[2]{%
  \rowvectoks={#2}\count255=#1\relax
  \advance\count255 by -1
  \rowvecnexta}
\newcommand{\rowvecnexta}{%
  \ifnum\count255>0
  \expandafter\rowvecnextb
  \else
  \begin{pmatrix}\the\rowvectoks\end{pmatrix}
  \fi}
\newcommand\rowvecnextb[1]{%
  \rowvectoks=\expandafter{\the\rowvectoks&#1}%
  \advance\count255 by -1
  \rowvecnexta
}

\def\bcen{\begin{center}}
\def\ecen{\end{center}}

\def\a{\alpha}

\def\p{\pi}

\def\G{\Gamma}

\def\PP{{\cal P}}

\def\AAA{{\cal A}}
\def\GG{{\cal G}}

\def\RRR{\mathbb{R}}
\def\CCC{\mathbb{C}}

\def\ZZZ{\mathbb{Z}}

\usepackage{bbold}
\def\11{\mathbb{1}}
\def\00{\mathbf{0}}

\def\eg{\mbox{\it e.g.\ }}
\def\ie{\mbox{\it i.e.\ }}
\def\=={\equiv}

\def\qed{\raise1pt\hbox{\vrule height5pt width5pt depth0pt}}
\def\iome{i\omega_n}

\def\cG0{{\cal G}_0}
\def\cG{{\cal G}}

\def\spinup{\uparrow}
\def\spindown{\downarrow}

\def\up{\uparrow}

\def\dw{\downarrow}

\def\ka{{\bf k}}
\def\ia{{\bf i}}

 \def\Im{\mbox{Im}}
\def\ie{\hbox{\rm i.e.\ }}
\def\eg{\mbox{\rm e.g.\ }}
\def\ie{\mbox{\rm i.e.\ }}
\def\=={\equiv}

\def\Im{{\rm Im}}
\def\Re{{\rm Re}}
\def\Tr{{\rm Tr}\,}
\def\det{{\rm det}\,}
\def\ep0{\epsilon_{p}}
\def\ed0{\epsilon_{d}}

\def\bhzrowvec{\rowvec{4}{c_{1\ka\spinup}}{c_{2\ka\spinup}}{c_{1\ka\spindown}}{c_{2\ka\spindown}}}

\begin{document}
\title{Strong Correlation Effects on Topological Quantum Phase Transitions in Three Dimensions}

\author{A.~Amaricci$^\mathrm{1}$}
\author{J.~C.~Budich$^\mathrm{2,3}$}
\author{M.~Capone$^\mathrm{1}$}
\author{B.~Trauzettel$^\mathrm{4}$}
\author{G.~Sangiovanni$^\mathrm{4}$}

\affiliation{$^\mathrm{1}$Democritos National Simulation Center,
Consiglio Nazionale delle Ricerche,
Istituto Officina dei Materiali (IOM) and
Scuola Internazionale Superiore di Studi Avanzati (SISSA),
Via Bonomea 265, 34136 Trieste, Italy}
\affiliation{$^\mathrm{2}$Institute for Theoretical Physics,
University of Innsbruck, 6020 Innsbruck, Austria}
\affiliation{$^\mathrm{3}$ Institute for Quantum Optics and Quantum
  Information, Austrian Academy of Sciences, 6020 Innsbruck, Austria}
\affiliation{$^\mathrm{4}$Institut f\"ur Theoretische Physik und
  Astrophysik, Universit\"at W\"urzburg, Am Hubland, D-97074 W\"urzburg, Germany}

\date{\today}

\begin{abstract}
We investigate the role of short-ranged
electron-electron interactions in a paradigmatic
model of three dimensional topological insulators, using dynamical
mean-field theory and focusing on
non magnetically ordered solutions. The non-interacting band-structure is
controlled by a mass term $M$, whose value discriminates between three
different insulating phases, a trivial band insulator and two
distinct topologically non-trivial phases.  We characterize the evolution of the
transitions between the different phases as a function of
the local Coulomb repulsion $U$ and find a remarkable dependence
of the $U$-$M$ phase diagram on the value of the local
Hund's exchange coupling $J$.  However, regardless the value of $J$, following the evolution of the topological transition line
between a trivial band insulator and a topological insulator, we find a critical value of $U$ separating a continuous transition from a first-order
one. When the Hund's coupling is significant, a Mott insulator is
stabilized at large $U$. In proximity of the Mott transition we
observe the emergence of an anomalous ``Mott-like'' strong topological
insulator state.
\end{abstract}


\maketitle

\section{Introduction}\label{SecIntro}

Recently, the conventional Landau classification of
matter has been complemented by the concept of topological
phases. Rather than being distinguished by the value of local order
parameters reflecting spontaneously broken symmetries, topological
phases of matter are characterized by global invariants describing
topological properties of the many-body wave-function. In the context of band structure physics, the  systematic search for topological
phases resulted in the discovery of a periodic table \cite{Schnyder2008PRB,Kitaev2009ACP} that builds upon the Altland-Zirnbauer symmetry classification \cite{Altland1997PRB}.
 
In two spatial dimensions (2D), the seminal theoretical
\cite{Kane2005PRL,Kane2005PRLa,Bernevig2006S} and experimental
\cite{Konig2007S} discovery of time reversal symmetry (TRS) protected
topological insulators (TIs) has inspired the theoretical community to
analyze the influence of Coulomb repulsion on such systems. Evidently,
the interplay between significant Coulomb interactions and topological
aspects of matter
holds the promise of nontrivial and intriguing effects emerging from the
competition between the localization tendency induced by strong correlations
and the peculiar band-structure of TIs. Regarding the influence of electron-electron interactions on 2D TIs, 
a wealth of results has been accumulated focusing
on several model Hamiltonians with different techniques (for a recent review see e.g. Ref.~\onlinecite{Hohenadler2013JOPCM}).
Among others, the effect of strong correlations on quantum spin Hall
insulators and the related topological transitions have been addressed using quantum Monte-Carlo calculations \cite{xLang2013PRB,xHung2013PRB,xHung2014PRB}, various cluster based
approaches \cite{Budich2012PRB,xGrandi2015PRB,xGrandi2015NJP,xWu2016PRB,xLaubach2014PRB}, 
as well as dynamical mean field theory (DMFT) and extensions
\cite{xRachel2010PRB,xChen2015PRB,Chen2015PRB,xWu2012PRB,Yoshida2012PRB,Yoshida2013PRB,Budich2013PRB,xNourafkan2014PRB,Amaricci2015PRL}.
Interestingly, the possibility of inducing the formation of a
topological phase by means of electronic repulsion has also been proposed
\cite{Raghu2008PRL,Ruegg2012PRL}.

In three spatial dimensions (3D), symmetry-protected TIs have been
have been suggested
theoretically~\cite{Fu2007PRL,MoorePRB2007,RoyPRB2009,Moore2010N} only
about two years later than their counterparts in 2D. Remarkably, 3D
TIs that preserve TRS exhibit non-degenerate Dirac
fermion surface states. This feature makes them particularly appealing
for fundamental science questions related to relativistic quantum
mechanics. Soon after their theoretical prediction, these topological
phases of matter have been experimentally detected in
three-dimensional
calchogenides~\cite{HsiehN2008,HsiehS2009,Zhang2009NP,ChenS2009,Wray2010NP}.


The interplay of strong electronic correlations with the various
topological phases, realizable in 3D \cite{Fu2007PRL}, pledges to unlock access to a
large variety of possible anomalous states. For instance, some Iridate compounds, with a pronounced three-dimensional character and the concomitant presence of strong spin-orbit coupling and
sizeable electron-electron interaction, have been proposed either as
correlated 3D TIs~\cite{Pesin2010NP} or to host even more exotic,
fractionalized topological states~\cite{Maciejko2014PRL,xMaciejko2015NP}.
The role of interactions in enhancing the effect of spin-orbit coupling
has been proposed as key to understand the onset of a topological
Kondo insulating state in both SmB$_6$  and
PuB$_6$~\cite{Dzero2010PRL,Dzero2012PRB,Zhang2013PRX,xLu2013PRL,Deng2013PRL}.
Additionally, an interaction driven 3D topological Mott insulating
state has been discussed in Refs.~\onlinecite{Zhang2009PRB,Herbut2014PRL},
while the existence of a U(1) spin-liquid with non-trivial spinon gap
has been proposed in Ref.\onlinecite{Pesin2010NP} for a class of pyrochlore Iridates.
More recently a Platinum oxide, Ca$_2$PtO$_4$, has also been suggested
to become a $d$-electron weak TI if doped with holes
~\cite{Li2015PRB}. Moreover, the realization of 3D TIs in oxides\cite{Kargarian2013PRL}, which have
typically sizeable values of the interaction parameters, gives a further
motivation to study the interplay of topology and electronic interactions.
However, the more complicated topological structure of the 3D TIs with
respect the their 2D counterparts has not yet allowed for a systematic
study of correlated effects in 3D.

In this work, we present a thorough study of the effects of a local
electronic interaction on the properties of a paradigmatic model of a 3D TI.
Using DMFT, we solve non-perturbatively a two-band Hubbard
model in presence of a multi-orbital density-density interaction,
featuring a Hund's coupling term.
We characterize the evolution of the non-trivial topological
states of the system as a function of the interaction strength.
We obtain the phase-diagrams of the model in the two-dimensional space
spanned by the Hubbard repulsion
and the crystal-field splitting, and determine the transition lines
between insulating phases with different topological properties.
One of the most prominent features of the model is that the transition line between
the trivial band insulator and the strong topological insulator is divided in two parts by a critical point.

For weak interactions the phase-transition is adiabatically connected with the non-interacting
transition and it is continuous. On the contrary, in the large interaction regime beyond the
critical point, the topological quantum phase
transition (TQPT) becomes of first-order. Unlike the conventional
scenario, the TQPT takes place without continuous closing of the
spectral gap, which abruptly ``inverts'' across the
transition.
Interestingly, this result extends to the 3D case
a recent finding of ours for quantum spin Hall insulators in 2D~\cite{Amaricci2015PRL}.
Reducing the crystal-field splitting, the strong topological
insulator, in the present case, turns into a weak topological
insulator through a transition which remains continuous, irrespective
of the strength of the interaction. This effect has no analog in the
previously studied 2D model.

This sequence of transitions occurs for every value of the Hund's
exchange coupling $J$, even if the shape of the phase diagram in the
$U$-$M$ plane is strongly influenced by the presence of the exchange
coupling.
A qualitative effect of $J$ is that it allows for a topologically
trivial Mott insulating state for large values of the interaction
strength. This leads to one further TQPT which separates the weak 3D
TI from the trivial Mott Insulator (MI).
We finally point out that, in a tiny slice bordering the transition
line between the 3D TI and MI, we observe the appearance of an anomalous
strong ``Mott-like topological insulator''. Such anomalous strong TI
is continuously connected to the MI but separated by a first-order
discontinuity from the weak 3D TI state.

The article is organized as follows. In Sec.~\ref{SecModel}, we
introduce the non-interacting model for the 3D TI and discuss its
solution within the framework of DMFT. We also briefly describe the
calculation of the $\ZZZ_2$ topological invariant with and without
interactions.
The multiple phases of the non-interacting model are reviewed in Sec.~\ref{secNonInteracting}.
The effects of the electronic correlations on the model solution are
discussed in Sec.~\ref{secPhaseDiagram}, where we also present the
phase diagrams of the full model.
In Sec.~\ref{secTQPT}, we illustrate the ways the different topological
phase transitions occur.
Finally, the transition to the Mott insulator is described in
Sec.~\ref{subsecMottTransition} which precedes the conclusions of the
paper.

\section{Model}\label{SecModel}
We consider a two-orbital Fermi-Hubbard model defined on a
three-dimensional cubic lattice. The model Hamiltonian reads:
\eq{
H = \sum_\ka\Psi^\dag_\ka \hat{\bf H}(\ka)\Psi_\ka + H_\mathrm{int}
}{model_hamiltonian}
where the spinor $\Psi_\ka\! =\! \bhzrowvec$ collects all the operators $c_{\ka l\sigma}$
($c^\dag_{\ka l\sigma}$), which annihilate (create) an electron at the
orbital $l\!=\!1,2$ with momentum $\ka$ and spin $\sigma$.
In order to write the explicit expression for $\hat{\bf H}(\ka)$
let us introduce the 4$\times$4 $\Gamma$-matrices, defined as follows:
\begin{equation}
  \begin{split}
    \Gamma_0 &= \11 \otimes \11 = \begin{bmatrix}\11&\00\\ \00&\11\end{bmatrix}\\
    \Gamma_1 &= \sigma_z \otimes \tau_x = \begin{bmatrix}\tau_x&\00\\ \00&-\tau_x\end{bmatrix}\\
    \Gamma_2 &=-\11 \otimes \tau_y = \begin{bmatrix}-\tau_y&\00\\ \00&-\tau_y\end{bmatrix}\\
    \Gamma_3 &= \sigma_x \otimes \tau_x = \begin{bmatrix}\00&\tau_x\\ \tau_x&\00\end{bmatrix}\\
    \Gamma_4 &= \sigma_y \otimes \tau_x = \begin{bmatrix}\00&-i\tau_x\\ i\tau_x&\00\end{bmatrix}\\
    \Gamma_5 &= \11 \otimes \tau_z = \begin{bmatrix}\tau_z&\00\\ \00&\tau_z\end{bmatrix}\\
\end{split}
\label{gamma_matrices}
\end{equation}
where $\tau_{x,y,z}$ and $\sigma_{x,y,z}$ are the Pauli matrices respectively in orbital and in spin space and $\11$
is the unit matrix. In terms of these $\Gamma_{\mathrm i}$ matrices we
have:
\eqsplit{
\hat{\bf H}(\ka)=
 M(\ka)\Gamma_5 + & \lambda \sin(k_{\mathrm x})\Gamma_{\mathrm 1} + \cr
& \lambda \sin(k_{\mathrm y})\Gamma_{\mathrm 2} + \lambda \sin(k_{\mathrm z})\Gamma_{\mathrm 3}
}{model_H0}
with $M(\ka)\!=\!M\!-\!\epsilon[\cos(k_x)\!+\!\cos(k_y)\!+\!\cos(k_z)]$.
This term of the model Hamiltonian would describe a system of two bands
of width $W\!=\!6\epsilon$, hybridizing with an amplitude $\lambda$ and
separated by a crystal-field splitting $2M$. In the following we
shall set $\epsilon$ as our energy unit.
In addition we fix $\lambda\!=\!0.3$ for definiteness having checked that qualitatively similar
results can be obtained for different values of $\lambda$.
Throughout this paper we consider a total density of two electrons per
site, which corresponds to a global half-filling of our band-structure
and leads to a particle-hole symmetry.

The second term of the model Hamiltonian (\ref{model_hamiltonian})
describes the screened Coulomb repulsion. We assume a local interaction
with a full orbital structure, namely {\it  inter}- and {\it intra}-orbital repulsion and
the  Hund's coupling $J$. This describes the exchange effect favoring 
high-spin configurations, which correspond to the two electrons occupying
different orbitals. More explicitly the interaction is:
\eqsplit{
H_\mathrm{int} = &U\sum_{\ia\,l}n_{\ia l\up}n_{\ia l\dw} +
(U-2J)\sum_{\ia\, l\neq
  l'}n_{\ia l\up}n_{\ia l'\dw}\cr
& + (U-3J)\sum_{\ia\, l\neq
  l'}\left(n_{\ia l\up}n_{\ia l'\up} + n_{\ia l\dw}n_{\ia l'\dw}\right)
}{model_interaction}
where $U$ is the strength of the electron-electron interaction and
$n_{\ia l\sigma}=c^\dag_{\ia l\sigma}c_{\ia l\sigma}$ is the local density for the orbital $l$
and spin $\sigma$ and $c_{\ia l\sigma}=2\p/V\sum_\ka e^{-i\ka\cdot\ia}c_{\ka l\sigma}$.
We notice that this Hamiltonian only contains the ``density-density" part of the Hund's
exchange and neglects the so-called pair-hopping and spin-flip terms \footnote{In Ref. \onlinecite{Budich2013PRB} the robustness of the topological transitions in the Bernevig-Hughes-Zhang-Hubbard model against the pair-hopping and spin-flip terms has been verified.}.

\begin{figure}
\includegraphics[width=0.35\textwidth]{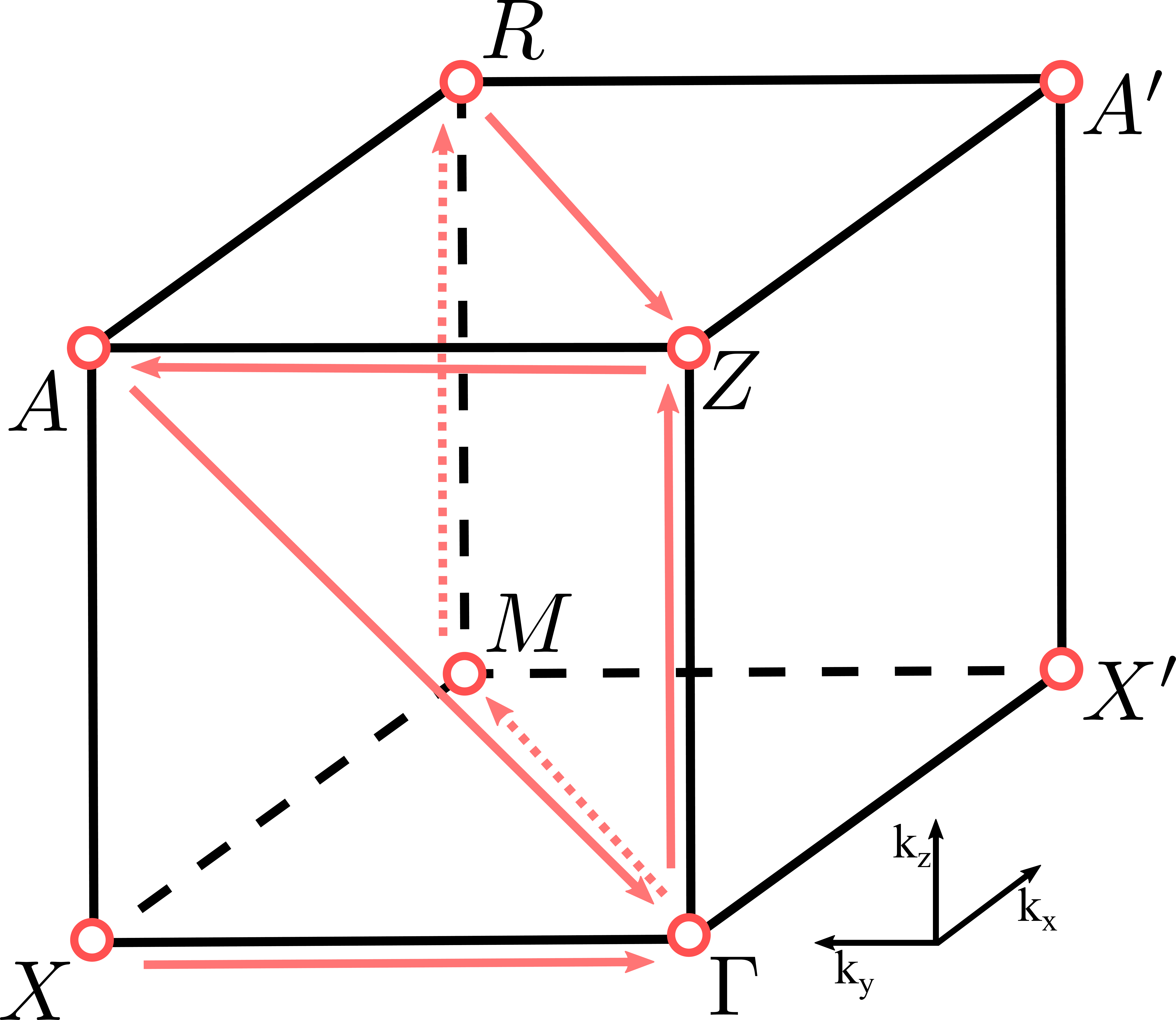}
\caption{(Color online) Unit cell of the cubic reciprocal lattice. The
  arrows indicate the three dimensional path in the reciprocal space
  used to depict the bands in the next \figu{fig2}
}
\label{fig1}
\end{figure}

\paragraph*{DMFT solution --}
We investigate the solution of the model (\ref{model_hamiltonian}) using
Dynamical Mean Field Theory (DMFT).
Within DMFT the quantum many-body lattice Hamiltonian
(\ref{model_hamiltonian}) is mapped onto an effective quantum impurity
problem. The impurity site features all the local interactions
of the original problem.
In our case, it corresponds to a two-orbital impurity with interactions of the same
form of Eq. (\ref{model_interaction}) coupled to an effective bath.
The bath is a frequency dependent quantity whose functional form is obtained
self-consistently by imposing that the impurity problem
reproduces the local physics of the lattice problem.
The DMFT mapping therefore approximates the
self-energy of the lattice problem with a momentum-independent
one obtained from the solution of the effective quantum impurity problem.
In terms of this function the self-consistency condition reads
\eqsplit{
\hat{\GG}^{-1}_0(z) & = \sum_\ka
\left[(z+\mu)\hat{\11}-\hat{\bf
    H}(\ka)-\hat{\bf\Sigma}(z)\right ]^{-1} +
\hat{\bf\Sigma}(z)\\
& = \hat{\bf G}^{-1}_\mathrm{loc}(z) + \hat{\bf\Sigma}(z)\\
}{WeissEq}
for $z\in\CCC$. Equation (\ref{WeissEq}) relates the {\it Weiss} Field
$\hat{\GG}^{-1}_0(z)$, describing the properties of the effective bath, to the
local physics of the lattice problem expressed by the local Green's
function $\hat{\bf G}_\mathrm{loc}(z)$.
All the effects of the interactions are included in the self-energy
$\hat{\bf\Sigma}(z)$ which is a 4$\times$4 matrix.
Owing to the symmetries of our model, including the particle-hole symmetry
at half-filling, the self-energy becomes diagonal in the spin-orbital basis, and it turns out to
acquire the following structure in terms of the $\Gamma$-matrices:
\eq{
\hat{\bf \Sigma}(z) =
\sum_{k\!>\!0} g_k z^{2k} \Gamma_5 +
\sum_{k\!\geq\!0} u_k z^{2k+1} \Gamma_0
}{Self}
with $g_k, u_k\in\RRR$.
In particular, the real- and imaginary-part of the Matsubara
self-energy $\hat{\bf \Sigma}(\iome)$ satisfies the relation:
\eq{
\hat{\bf  \Sigma}(\iome) = \Re{\Sigma(\iome)}\Gamma_5 + i\Im{\Sigma(\iome)}\Gamma_0
}{MatsubaraSelf}
which means that the full self-energy is parametrised by a single scalar complex function $\Sigma(\iome)$.
Eq. (\ref{MatsubaraSelf}) implies that the imaginary part of the
self-energy is the same for all the spin and orbital components, while
the real part is the same for the two spin components and it has the
opposite sign for the two orbitals. As a matter of fact the real part
of the scalar self-energy is simply added to the bare splitting
$M$. Therefore we can define an {\it{effective mass term}} in terms of
the low-frequency limit of the self-energy \cite{PoterPaperI,Parragh2013PRB,Budich2013PRB}
\eq{
M_\mathrm{eff} = M + \Re{\Sigma(\omega=0)}
}{m_effective}
In the following we will show that this quantity controls the location
of the topological phase transitions in the presence of interactions.

We solve the effective problem by using an exact diagonalization impurity
solver~\cite{Georges1996RMP,Capone2007PRB,Weber2012PRB}.
In this scheme the effective bath is discretized to a finite number 
$N_b$ of levels coupled to the $N_o=2$ impurity orbitals.
The resulting Hamiltonian is solved by means of the Lanczos technique
which allows to determine the lowest part of the spectrum as well as the
dynamical correlation functions at zero and low temperatures.
In this work we performed calculations for $N_b\!=\!8$ and checked the
convergence as a function of $N_b$ by comparing with $N_b\!=\!10$ for selected values of
the model parameters.

\begin{figure}
  \includegraphics[width=0.45\textwidth]{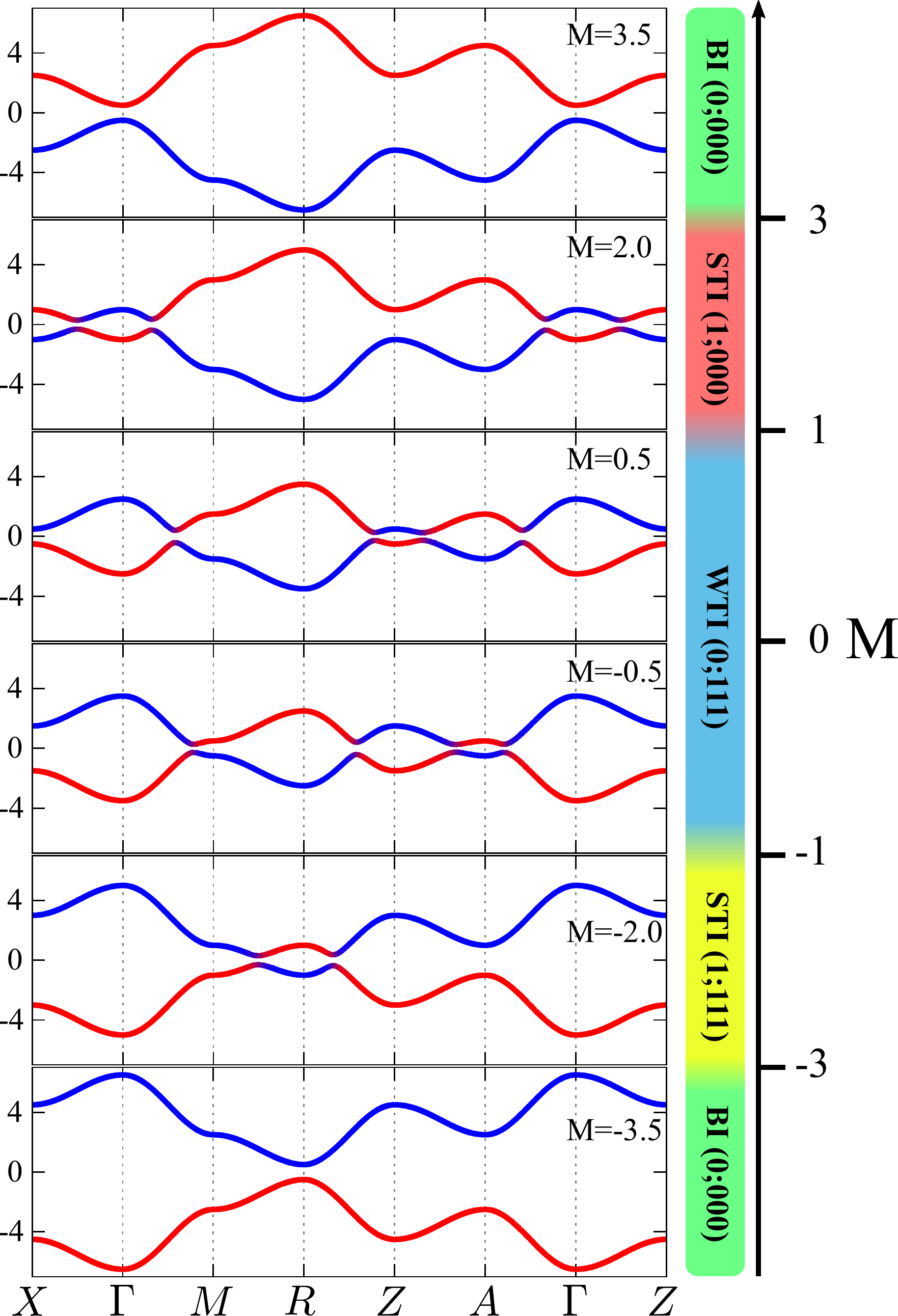}
  \caption{
    (Color online) Evolution of the non-interacting band-structure (left) as a function of $M$ and consequent ``phase diagram" (right) of  the model (\ref{model_H0}).
    The coloured intervals on the right indicates the different topological phases of the system.
    The colors of the band structures correspond to the orbital
    character. The plot also illustrates the difference between the strong topological
    insulator realized at $M>0$, \ie STI$_\G$, and the one realized at
    $M<0$, \ie STI$_R$.
  }
  \label{fig2}
\end{figure}

\begin{figure*}
  \includegraphics[width=0.85\textwidth]{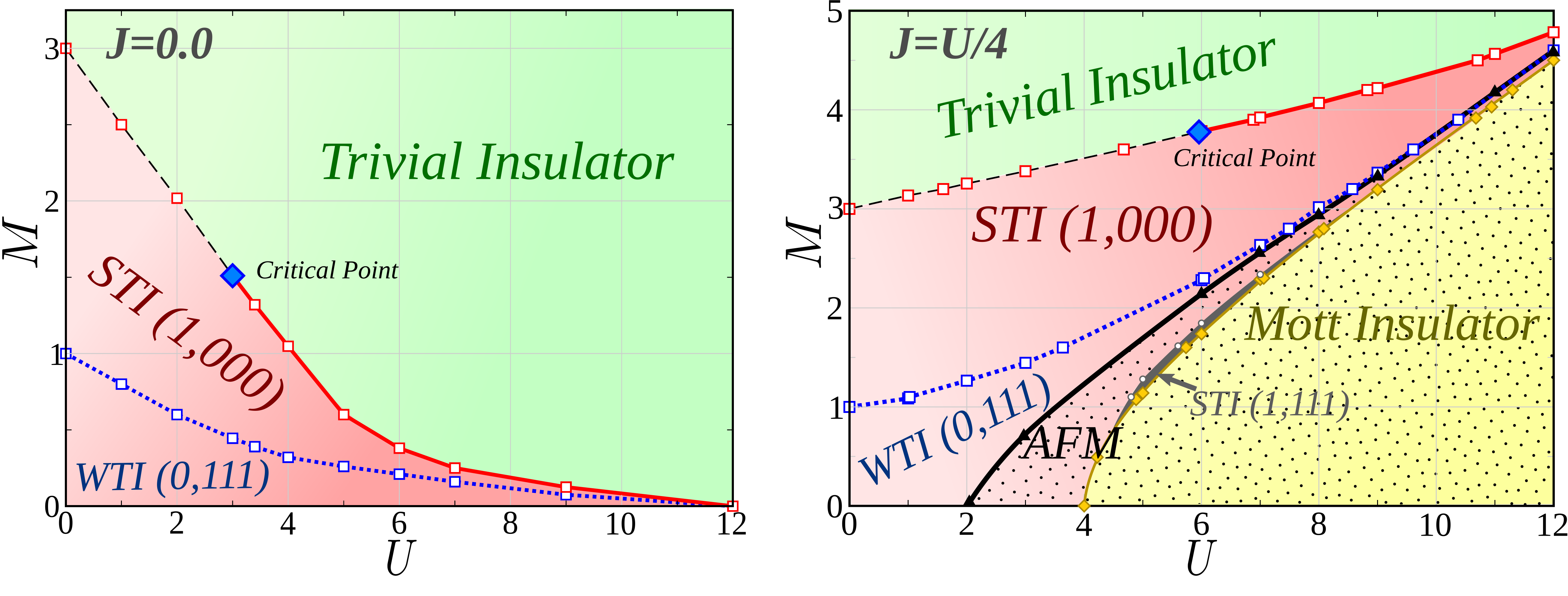}
  \caption{(Color Online) Interacting phase diagram of the three-dimensional model
    (\ref{model_hamiltonian}) as a function of $M$ and $U$, with $J=0$ (left panel) or $J=U/4>0$ (right
    panel). Data are for $\lambda=0.3$. The solid (red) line ends in
    the Critical Point and is continued beyond this value by the
    dashed (orange) line. The union of these two lines indicates the
    boundary between the Trivial Insulator and the Strong Topological
    Insulator $(1,000)$. The dotted (blue) line indicates the transition
    from the STI to the weak topological insulator $(0,111)$. In the
    right panel the solid (yellow) line is the boundary of the Mott
    Insulating region. The filled (grey) area near the Mott phase shows
    the re-entrant anomalous STI $(1,111)$ phase, precursor of the
    Mott transition.
    The dotted area in the right panel indicates the AFM phase
    region. The solid (black) line corresponds to the boundary of the
    AFM phase, which is separated from the non-trivial region by a
    first-order transition. 
  }
  \label{figPD}
\end{figure*}

\paragraph*{Topological invariants -- }\label{secTopInvariants}
Three dimensional topological insulators preserving TRS
with $\mathcal T^2=-1$, are classified
by four $\ZZZ_2$ topological invariants~\cite{Fu2007PRL}, here denoted
by $\vec{\nu}=(\nu_0;\nu_1\,\nu_2\,\nu_3)$. The latter three numbers
$\nu_i,~i=1\ldots 3$ are so called weak topological invariants that
describe stacks of 2D topological insulators in the same symmetry
class. Such weak topological insulators are only defined in lattice
systems. By contrast, the strong topological invariant $\nu_0$ also
exists in continuum systems and exhibits a much stronger robustness
against disorder. In the presence of interactions, all these
invariants can be generalized to topological properties of the single
particle Green's function
~\cite{Redlich1984PRD,So1985POTP,Ishikawa1987NPB,Volovik1988J,Gurarie2011PRB,He2015AE1,He2015AE2}.

As it has been shown in Ref. \onlinecite{Wang2012PRX}, the practical
computation of such interacting invariants formally reduces to that of
a non-interacting problem, where the role of the Bloch Hamiltonian is
played by the so-called {\it topological Hamiltonian} which reads as
\eq{
\hat{\mathbf{H}}_\mathrm{t}(\ka)
=-\hat{\mathbf{G}}^{-1}(\ka,\omega\!=\!0)
=\hat{\mathbf{H}}(\ka)+\hat{\mathbf{\Sigma}}(\omega\!=\!0).
}{topological_hamiltonian}
For the insulating phases considered in this work, the imaginary
part of the self-energy is linearly vanishing in $\omega\!=\!0$. The
only finite contribution thus comes from the real-part
$\Re{\hat{\mathbf{\Sigma}}(\omega\rightarrow0)}$.

If  inversion symmetry is not broken, as for our model
(\ref{model_hamiltonian}), the definition of the topological
invariants is drastically simplified
~\cite{Fu2007PRB,Wang2012PRB}. Following
Ref. \onlinecite{Wang2012PRB}, we obtain the eigenstates of
$\hat{\mathbf{H}}_\mathrm{t}$ at the eight time reversal invariant
momenta (TRIM) $K_{i\!=\!1\dots 8}$ and for the occupied bands
$\a$. These states can be chosen as eigenstates of the parity operator
$\PP$ with eigenvalue
$p_\mathrm{\tiny i\a}=\pm 1$. The different $\ZZZ_2$ invariants
$\nu_{c=0,\dots,3}$ are then calculated using
the relation:
\eq{
(-1)^{\nu_c} = \prod_\mathrm{\tiny i\in\AAA_c,\, \a}\sqrt{p_\mathrm{\tiny i\a}} .
}
where the phase convention $\sqrt{-1}=+i$ is used, $c=0,\dots,3$
labels the components of the global invariant vector
$\vec{\nu}$, and $\AAA_c$ indicates the set of TRIM points as follows
(see Fig. \ref{fig1} for the labels of the TRIM):
\eqsplit{
\AAA_0 & = \left[\Gamma, X' , M , X , A , Z , A' , R \right]\\
\AAA_1 & = \left[X' , M , R , A' \right]\\
\AAA_2 & = \left[X , A , R , M \right]\\
\AAA_3 & = \left[Z , A' , R , A \right],\\
}{trim_sets}
where the location in momentum space of the TRIM points is depicted in
\figu{fig1}. Quite intuitively, the sets $\AAA_c, c=0\ldots 3$ entering the weak
topological invariants are confined to 2D planes with three orthogonal
orientations in the 3D Brillouin zone while the strong topological
invariant involves all eight TRIM.

\section{Topological Transitions in the non-interacting limit}\label{secNonInteracting}
We begin our analysis from the non-interacting model $U\!=\!J\!=\!0$.
The Hamiltonian (\ref{model_H0}) describes a
three-dimensional system undergoing a series of topological quantum
phase transitions as a function of the crystal-field splitting term $M$.
The band structure along the path in the reciprocal
space depicted in \figu{fig1} for the various phases is reported in
\figu{fig2}.
For $|M|<1$ the model describes a weak topological insulator (WTI)
with a global invariant $\vec{\nu}\!=\!(0;111)$.

The smallest direct gap is placed near the point $Z$ ($A$)
for $M>0$ ($M<0$) in correspondence to a slight inversion of the band
character. A more robust band inversion takes place close to the
$\Gamma$ point. A Dirac-like gap closure takes place at the points $Z$
($A$) for $M=1$ ($M=-1$).

For values of $M$ in the range $1\!<\!|M|\!<\!3$ there is a
transition to a strong topological insulating (STI) state.
In the $M\!>\!0$ region such state is characterized by a global invariant
$\vec{\nu}\!=\!(1;000)$, while for $M\!<\!0$ the STI has
$\vec{\nu}\!=\!(1;111)$~\cite{SbierskiPRB2016}.
The values of the topological invariants reflect
differences in the band structures: For $M\!>\!3$ the smallest
direct gap and band inversion takes place around the $\Gamma$ point,
while for $M\!<\!-3$ these are located in proximity of the $R$ point.

For $|M|\!=\!3$ the system undergoes a TQPT from the STI to the trivial
band insulator (BI) at $|M|\!>\!3$, see \figu{fig2}.
At the transition the topological invariant undergoes a sudden change to the
trivial value $\vec{\nu}\!=\!(0;000)$.
The TQPT is continuous with the formation of a Dirac-like gap closure at
the point $\Gamma$ ($M\!=\!3$) or $R$ ($M\!=\! -3$). For this reason
we will indicate with $STI_\Gamma$ and $STI_R$ respectively the strong
topological phases for $M\!>\!0$ and $M\!<\!0$.

\section{Interacting Phase Diagrams}\label{secPhaseDiagram}
We now turn our attention to the effects of the interaction on the
properties of the TQPT in the model (\ref{model_hamiltonian}).
For definiteness we restrict our analysis to the positive values of
the splitting $M$. Analogous results can be obtained
for the $M<0$ case.

Our results are summarized in the phase-diagrams in the $U$-$M$ plane reported in
\figu{figPD}. The figure compares the phase-diagram in the $U$-$M$
plane for zero Hund's coupling $J=0$ (left panel) to that for $J=U/4$ (right panel).
In both diagrams one clearly sees that the two transitions of the
non-interacting system occurring from $M=3$ (between BI and STI) and
$M=1$ (between STI and WTI) are continued into boundary lines that
extend in the diagram as $U$ is increased. The phase boundaries have
been drawn according to explicit evaluation of the topological
invariants from the single-particle Green's
function~\cite{Redlich1984PRD,So1985POTP,Ishikawa1987NPB,Volovik1988J,Gurarie2011PRB}.

However, at least two macroscopic differences appear immediately: (i)
a Mott insulating phase is obtained only for finite $J$ and (ii) the
phase boundaries separating respectively BI from STI and STI from WTI
have a completely different behavior as a function of $U$, where the critical
$M$ decreases as a function of $U$ in the absence of $J$, and it has
the opposite behavior for finite $J$.  We notice that for a
wide range of $J$ the phase diagram is similar to that of $J=U/4$, with the only exception of very
small values of $J/U$, where the boundary lines first decrease (as for $J=0$)
and then increase (as for $J=U/4$) as a function of $U$ \cite{Werner2007PRL}.

\begin{figure}
  \includegraphics[width=0.49\textwidth]{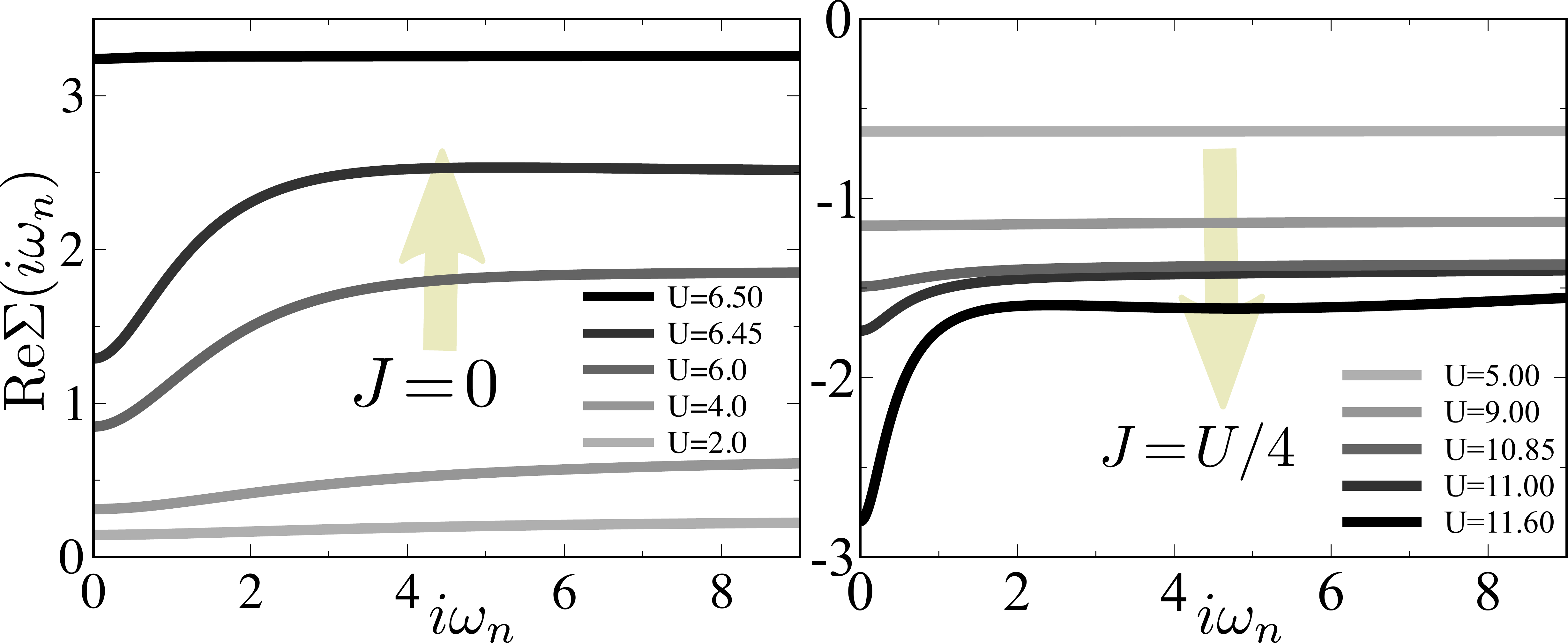}
  \caption{(Color Online) Real part of the scalar Matsubara
    self-energy $\Re\Sigma(i\omega_n)$ for an increasing
    interaction strength $U$.
    Data are shown for $M=0.3$ with $J=0$ (left panel), or $M=4.5$ with
    $J=U/4$ (right panel). The figure illustrates the opposite algebraic sign of the
    self-energy in the two investigated cases. It also points out the
    pronounced frequency dependence at low-energy.
  }
  \label{figS}
\end{figure}

We can understand these differences analyzing the real-part of the self-energy $\Re\hat{\bf \Sigma}$, which
renormalizes the bare splitting between the orbitals. 
In \figu{figS} we report the evolution of  $\Re\hat{\bf \Sigma}$ as $U$ is increased for
$J=0$ (left panel) and  $J>0$ (right panel).

The difference is apparent: for $J=0$ the self-energy is positive and increases with $U$, while for
$J>0$ it is a negative quantity whose absolute values grows with
$U$. This implies that in the absence of $J$ the interaction
effectively enhances the crystal-field splitting, while an opposite
effect takes place for finite $J$.
This is easily understood since   the Hund's coupling favours configurations
in which the electrons populate the different orbitals to maximize the spin (One can
visualize this very simply in the atomic limit where all the kinetic terms are neglected),
while in the absence of $J$ intra- and inter-orbital repulsion are identical and
the electrons can freely populate the orbitals.

As a result, for $J=0$ the interactions lead to a very large effective
splitting, which favours a BI configuration in which only one orbital
is populated. For this reason the boundary lines decrease as a
function of $U$.  On the contrary, for finite $J$ and large
interaction a weak effective splitting leads to a Mott insulator with
one electron per orbital and increases the threshold to reach the
topological transitions.

The presence of large local moments in the high-spin Mott insulating
region favours the onset of magnetic ordering. We verified that, as
expected, our model solution is unstable towards the formation of a
topologically trivial G-type antiferromagnetic (AFM) state for large
$U$. We computed the boundary of the resulting AFM region in the phase diagram
(see \figu{figPD}) when allowing for a broken symmetry solution. In
addition we found that the formation of the
AFM state takes place through a first-order transition, with the
discontinuous formation of a finite order parameter. 
Interestingly the AFM region does not hide the TQPT between BI and
STI$_\Gamma$. 

\begin{figure}
\includegraphics[width=0.45\textwidth]{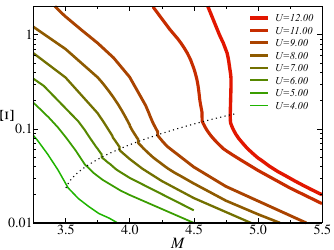}
\caption{(Color Online) Evolution of the correlation strength measured
  by $\Xi$  as a function of $M$ and for increasing values of $U$. All data are for
  $J=U/4$.
  The dotted line indicate the transition points. }
\label{figXi}
\end{figure}

In addition to the $J$-dependence, the evolution of the self-energy of
\figu{figS} reveals another important feature. In both cases for small
values of $U$ the real part of the self-energy is essentially constant
as a function of the imaginary frequency. This is consistent with a
Hartree-Fock solution (or static mean-field) which
can indeed obtained at lowest order in a diagrammatic expansion. As $U$ increases
$\Re\hat{\bf \Sigma}$ acquires a more and more pronounced frequency dependence. Roughly speaking
the dynamical nature of the interaction effects is measured by
the difference between the low-frequency limit, in which the dynamical correlation effects are dominant
and the large-frequency limit, that we label as a static mean-field value $\hat{\bf \Sigma}_\mathrm{MF}\!=\!\Re\hat{\bf
  \Sigma}(\iome\rightarrow\infty)$. 
In \figu{figXi} we show the behavior of the correlation strength
$\Xi$, which we define as: 
\eq{
  \Xi=\frac{\Tr\!
    \left[
      \Gamma_5\Re\hat{\bf \Sigma}(0)\!-\!
      \Gamma_5\hat{\bf \Sigma}_\mathrm{MF}\!
    \right]
    \!}
  {\Tr\!
    \left[
      \Gamma_5\hat{\bf\Sigma}_\mathrm{MF}\!
    \right]}\,, 
}{correlation_Xi}
as a function of $M$ and for different values of $U$ and $J=U/4$.

For a fixed value of the interaction $U$ the system is driven through a TQPT by decreasing
$M$ below a given critical value.
In the weakly interacting regime the degree of correlation $\Xi$
remains small and smooth across the TQPT (for smaller values of $U$ than those considered in the figure, $\Xi$ is even more structureless at the topological transition).
This behavior changes dramatically in the strongly interacting regime:
The curves of $\Xi$ at constant $U$ become divided into to distinct parts: the large-$M$ piece resembles the weak dependence of the small-$U$ region. For $M$ below the TQPT (marked in the figure by a dotted line) $\Xi$ gets instead very rapidly large indicating a strong degree of many-body character of the solutions.
This characteristic behaviour of $\Xi$ reflects a simple physical effect:
The fully polarized BI with two electrons in the lowest
orbital bands is essentially unaffected by the strength of the interaction. On the
other hand the topological STI$_\Gamma$ and WTI region do not have a
full orbital polarization (hence a higher degree of hybridization). As
such they react more vigorously to
the presence of interaction, evolving from a weakly perturbed regime
to a strongly correlated topological state, as evidenced by $\Xi$.

\section{Topological Quantum Phase-Transitions.}\label{secTQPT}



We now study in detail the effect of correlations on the two transitions connecting respectively
 the BI from the STI$_\Gamma$ and the latter from the WTI.

As we mentioned in Sec. \ref{secTopInvariants}, the transition lines
can be traced according to the evolution of
the effective mass term $M_\mathrm{eff}$, which is the only way in which the interactions influence
the value of the topological invariants.

\begin{figure}
  \includegraphics[width=0.451\textwidth]{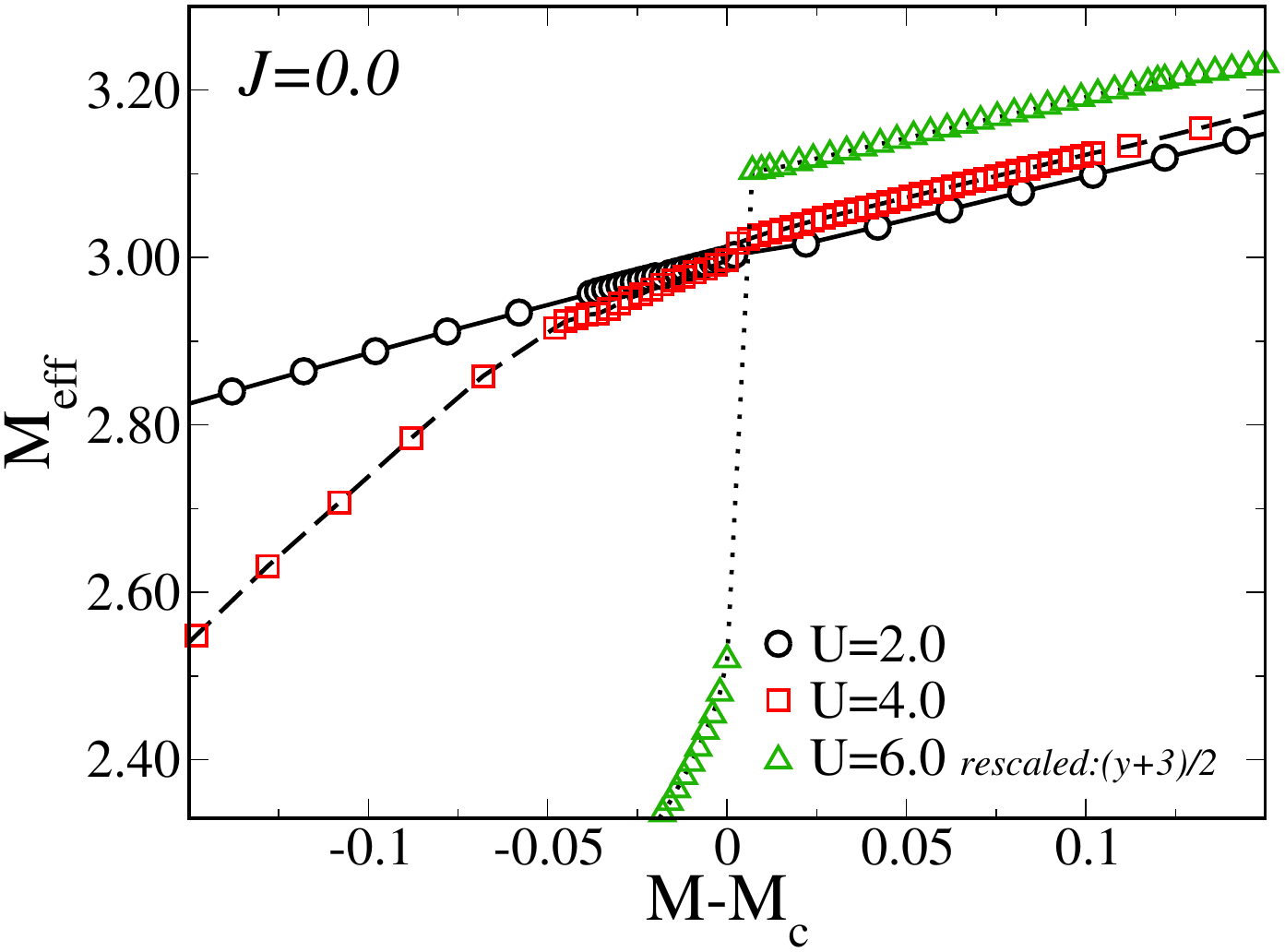}
  \includegraphics[width=0.451\textwidth]{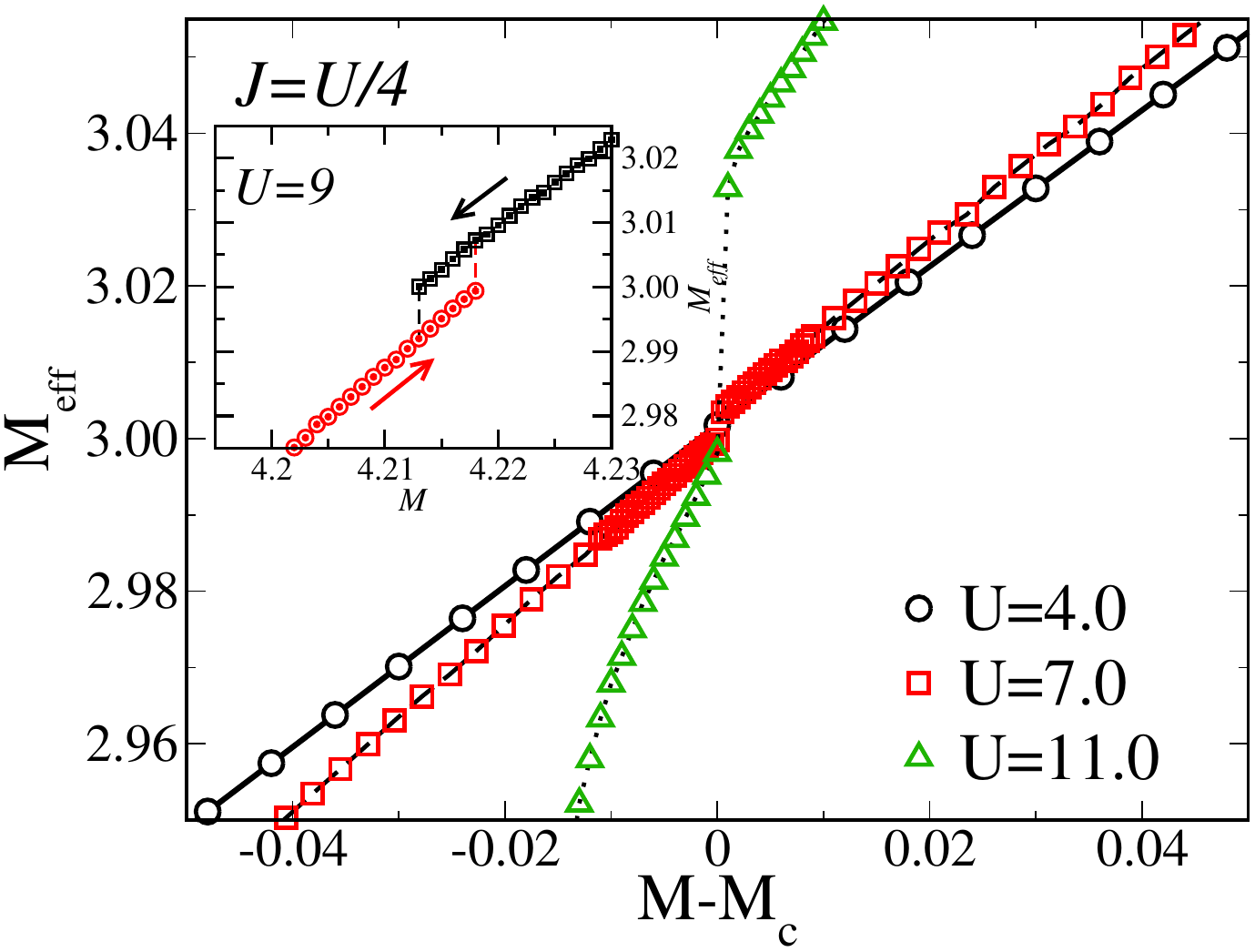}
\caption{(Color Online) Evolution of the effective Mass term
  $M_\mathrm{eff}$ as a function of the bare mass $M$. Panel (a)-(b)
  illustrates the transition from Trivial Band Insulator to Strong
  Topological Insulator for $J=0$(a) and $J>0)$(b).
  Inset: Hysteresis loop of $M_\mathrm{eff}$ for $U=9$ and $J=U/4$.
}
\label{figMeff}
\end{figure}

The TQPT from the trivial BI to STI$_\Gamma$ is
defined by the condition $M_\mathrm{eff}\!=\!3$ which generalizes the condition $M\!=\!3$ valid for the non-interacting case,
and naturally reduces to it for $U=0$ when the self-energy vanishes.
The whole $M_\mathrm{eff}\!=\!3$ line is characterized by a change
of the bulk topological invariant $\vec{\nu}$
from $(0,000)$ in the trivial BI phase to
$(1,000)$ in the STI$_\Gamma$.
However the nature of such transition changes dramatically from the
weak to the strong interaction regime, as evidenced by the behavior of
$M_\mathrm{eff}$ reported in \figu{figMeff} for
$J=0$ and $J>0$.
In both cases, for low values of the interaction $U$ the transition remains
continuous, similarly to the non-interacting regime discussed in
\secu{secNonInteracting}.
This reflects the fact that, if the correlation strength is small, the
TQPT can be very well described within a Hartree-Fock picture, as
already indicated by the behavior of $\Xi$.
In this effective single-particle picture a change in
the mass term $M$ perfectly compensates the effects of the interaction
$U$, ultimately leading to a continuous, non-interacting like,
TQPT. This behavior is illustrated in \figu{figMeff} for the smallest
values of $U$ for which the crossing of the $M_\mathrm{eff}=3$ line is
continuous.

By increasing the interaction strength $U$ above the critical value
$U>U_c$, the degree of correlation of the STI$_\Gamma$ is 
no longer negligible. Consequently, a renormalized single-particle description
of the TQPT breaks down.
In particular, the
strong dynamical dependence of the self-energy in this regime
does not allow to compensate the effects of the interactions by means
of a static change of $M$.
The transition is still positioned at $M_\mathrm{eff}=3$ but
the ground states in the trivial BI and in the more correlated
STI$_\Gamma$ can indeed no longer be continuously connected across the TQPT.
As a consequence, a first-order jump is required to move from one phase to the other.
The existence of such a first-order TQPT is demonstrated in
\figu{figMeff}, where we show the evolution of $M_\mathrm{eff}$ for
values of $U$ above the critical point. As the figure shows, in such regime the effective
mass term displays a discontinuity at $M_\mathrm{eff}=3$, irrespective of
the value of the Hund's coupling $J$.
As a further hallmark of the first-order
character of the correlated TQPT, we show the hysteretic
cycle across the transition in the inset to \figu{figMeff}.

This finding shows that electron-electron correlations profoundly change the nature
of the topological phase transitions, similarly to what we already found in two dimensions
~\cite{Amaricci2015PRL} for quantum spin Hall insulators.
In the three-dimensional system,  the transition from the BI to the
STI$_\Gamma$ is therefore continuous for small values of $U$ but it
becomes of first-order upon crossing the critical point, at which the orbital
fluctuations become critical~\cite{Amaricci2015PRL}.
The breakdown of the continuous character of the TQPT in Hamiltonian
models similar to our has been recently investigated by means of a
renormalization group analysis in Ref.~\onlinecite{Roy2015AE}.

Our findings resembles the marginal quantum criticality scenario proposed in
Refs.~\onlinecite{Imada2005PRB,Misawa2007PRB,Kurita2013PRB}, where a critical point
separates a first-order line from a second-order one.
However, a thorough analysis of the scaling properties near
the critical point are required to establish a more detailed connection, which is
however beyond the scope of this work.

Interestingly, while the increased level of correlation strength
dramatically affects the TQPT, it has a very weak impact on the
other topological transition, the one from STI$_\Gamma$ to the WTI. Here
the change of the bulk invariant from $\vec{\nu}=(1,000)$ to $\vec{\nu}=(0,111)$ is
always accompanied by a continuous evolution at $M_\mathrm{eff}=1$. To the best of
our accuracy we did not find any evidence of discontinuity, even at
the largest investigated value of the interaction strength $U$.
This behavior can be understood noting that, although
topologically distinct, the two non-trivial states react in a very
similar way to the large interaction having a comparable degree of orbital hybridization. Thus, unlike the BI to
STI$_\Gamma$ TQPT, in this case it is always possible to continuously
transform the STI$_\Gamma$ ground state into that of the WTI.


\begin{figure*}
\includegraphics[width=0.325\textwidth]{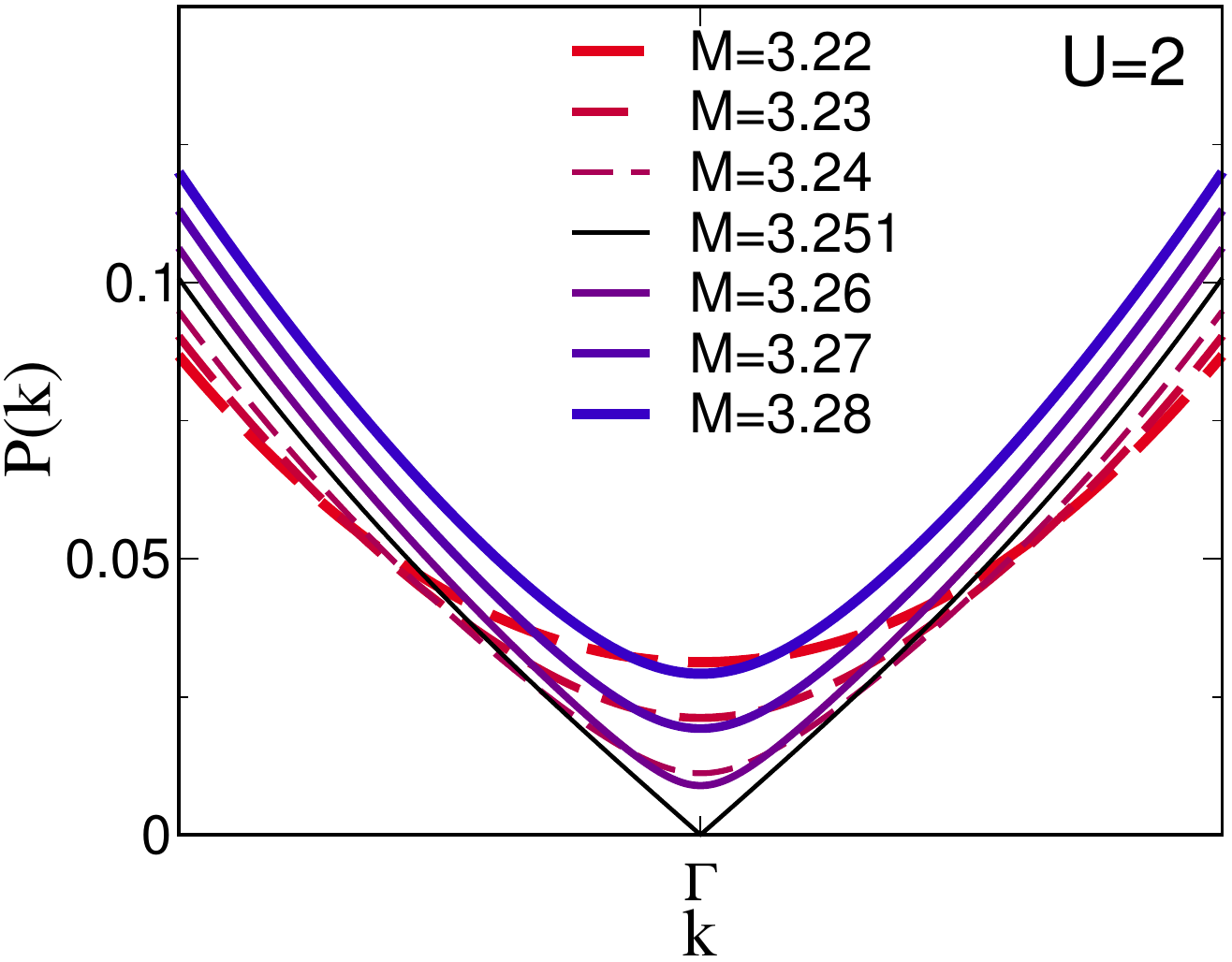}
\includegraphics[width=0.325\textwidth]{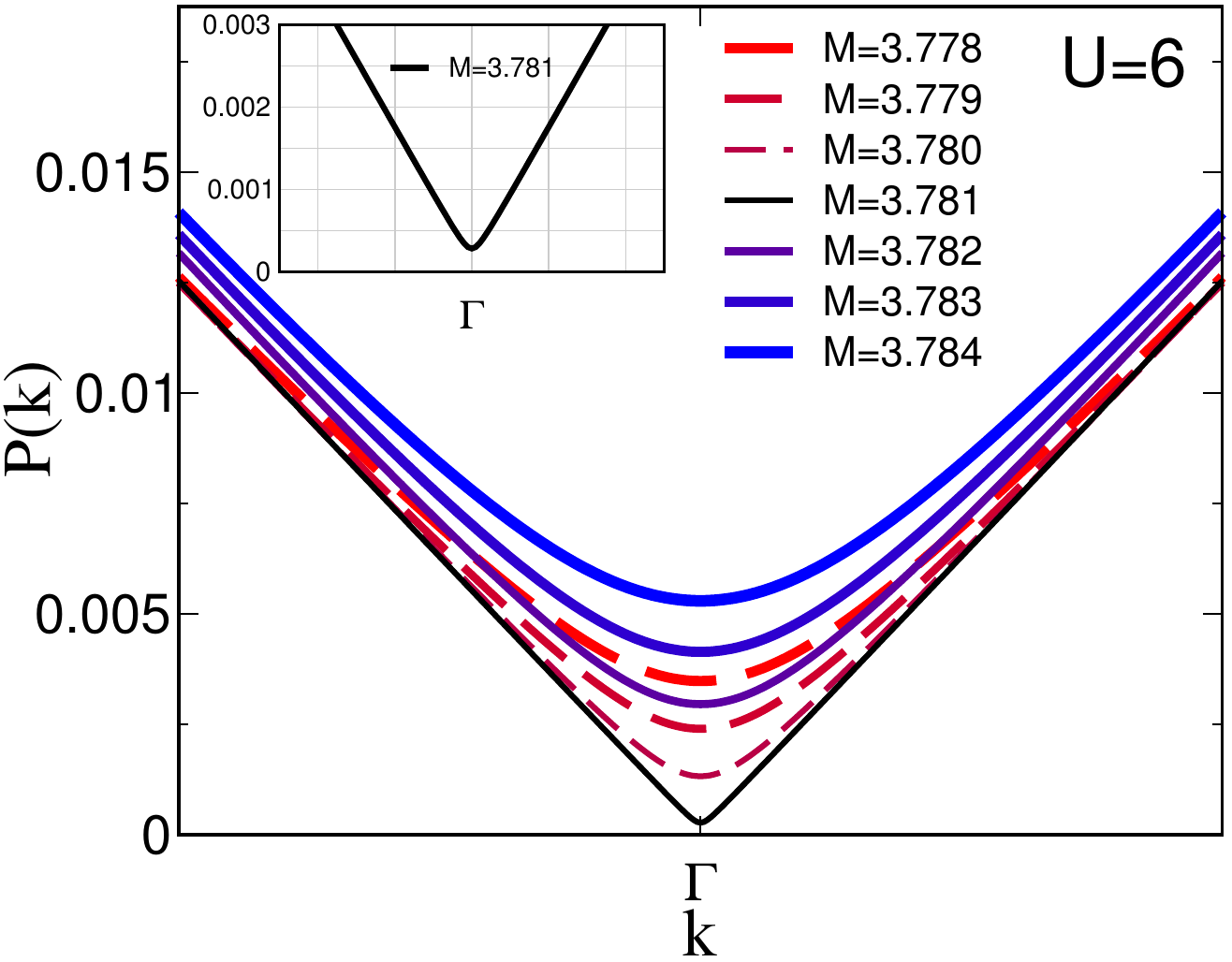}
\includegraphics[width=0.325\textwidth]{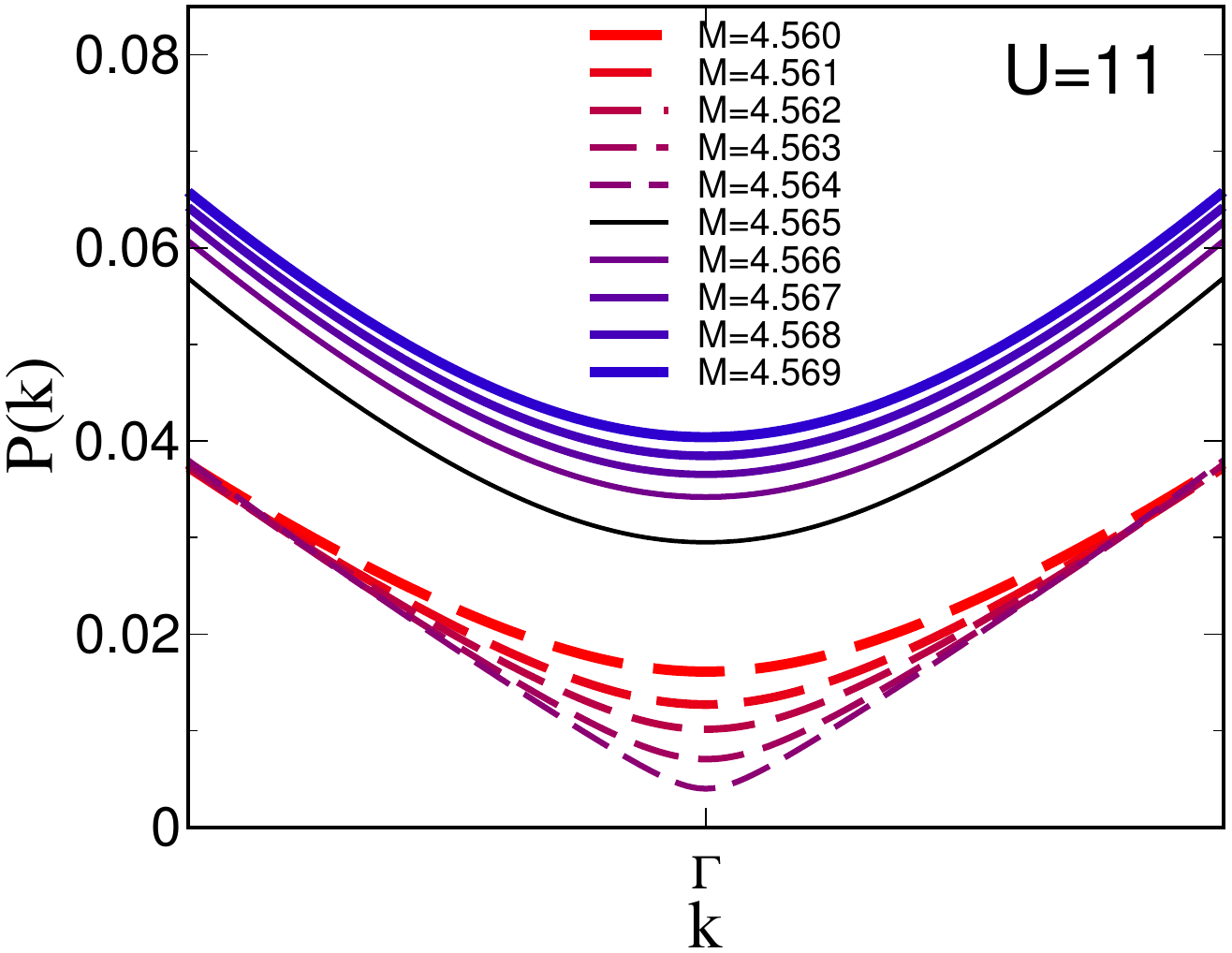}
\caption{
(Color Online) Absence of bulk gap closing: Evolution of the Green's functions poles, i.e.
the band dispersions. We find a continuous transition at weak interaction ($U=2$), contrasted
by the absence of bulk gap closure at strong interaction ($U=11$).
}
\label{figX}
\end{figure*}

\subsection{Absence of gap closing transition.}\label{subsecNoGapClosing}
As we discussed in the initial sections \ref{secTopInvariants} and \ref{secNonInteracting}, in the
non-interacting limit the TQPT are always continuous. This ultimately
depends on the fact that the band structure of the system evolves
smoothly with respect to changes in the model parameters, \eg
$M$, as long as the energy gap is preserved.
In presence of large interaction however this scenario is challenged
and a different behavior can be envisaged. We shall now show how the
change in the TQPT character across the phase-diagram affects the
spectral properties of the transitions.

Our results about the absence of a gap closing TQPT are summarized in \figu{figX}.
Each plot reports the behavior of the poles $P(\ka)$ of the Green's
function $\hat{\bf
  G}(\ka,\omega)\!=\!\left[\!(\omega\!+\!i\eta\!+\!\mu)\hat{\bf
    1}\!-\!\hat{\bf H}(\ka)\!-\!\hat{\bf \Sigma}(\omega)\! \right]^{-1}$, determined by the
condition $\det{\left[\hat{\bf G}(\ka,\omega)\right]}\!=\!0$, near the
point $\Gamma\!=\!(0,0,0)$ and along the path $[-R,\Gamma,R]$. The
curve $P(\ka)$ sets the position of the electronic excitations,
it determines the spectrum of the system and helps determining
the properties of the bulk spectral gap.
From left to right, the figure shows the evolution of the poles as a
function of $M$ across the BI-STI$_\Gamma$ transition for three values
of the interaction strength $U$: Below, at, and above the critical point.

For small values of $U$  the transition occurs through the formation of a
Dirac cone at the $\Gamma$ point, as expected by the
continuity of the TQPT with respect to the non-interacting case.
On the contrary, for $U>U_c$ the topological transition takes place
without a closure of the spectrum, as we show in the panels
(b)-(c).
Thus, at the strongly correlated transition point the gap jumps discontinuously
from the trivial to the topological value, although in proximity of
the critical point it becomes very small.
The absence of any gap-less state separating the two phases across the
TQPT is a remarkable effect induced by the strong electronic
correlation.

\section{Mott Transition.}\label{subsecMottTransition}
One of the qualitative changes brought in by a finite value of the
Hund's coupling is the onset of a Mott insulating region for large
$U$, as we reported in \figu{figPD} (right panel).
Indeed, in this case the system favours the formation of an orbital
un-polarized state, ultimately leading to an instability towards the
high-spin Mott phase as soon as the equal occupation of each orbital
is reached~\cite{Werner2007PRL,Amaricci2015PRL}.
Such a high-spin Mott insulator cannot be realized in the $J=0$ case,
where the strong interaction favours the formation of an orbitally
polarized configuration, \ie a trivial band insulator (see
\figu{figPD}).

In \figu{figMott}(a) we demonstrate the existence of a Mott transition
for large values of $U$.  The formation of a Mott insulating state is
signaled by the divergence of the imaginary part of the self-energy
at the chemical potential.  Thus we can identify the onset of the Mott
phase by means of the reduction to zero of the renormalization
constant $Z$,  defined as
$Z=\left[1-\partial{\Sigma}(\omega)/\partial\omega|_{\omega\rightarrow\mu}\right]^{-1}$.
The different phases of the system have an insulating
character, thus we can not interpret $Z$ in terms of a quasi-particle
weight.
As our results show, for a fixed value of the interaction $U$,
we can drive the $Z$ towards zero by decreasing the mass term $M$,
entering this way the high-spin Mott phase (see \figu{figPD}).
The Mott transition has a characteristic first-order behavior, with
all the observables showing a discontinuity at the transition
point and a small hysteretic behavior (not shown). The presence of
such a discontinuity is evident in the behavior of $Z$ for a small
enough value of $U$, but it becomes smaller approaching
the large $U$ regime.
The Mott transition line ends in a triple point, beyond which any
correlated topological state disappears, opening the way to a direct
transition from the polarized trivial BI to the high-spin Mott state~\cite{Werner2007PRL,Amaricci2015PRL}.

\subsection{Small region of anomalous STI$_R$ phase.}
\begin{figure}
  \includegraphics[width=0.49\textwidth]{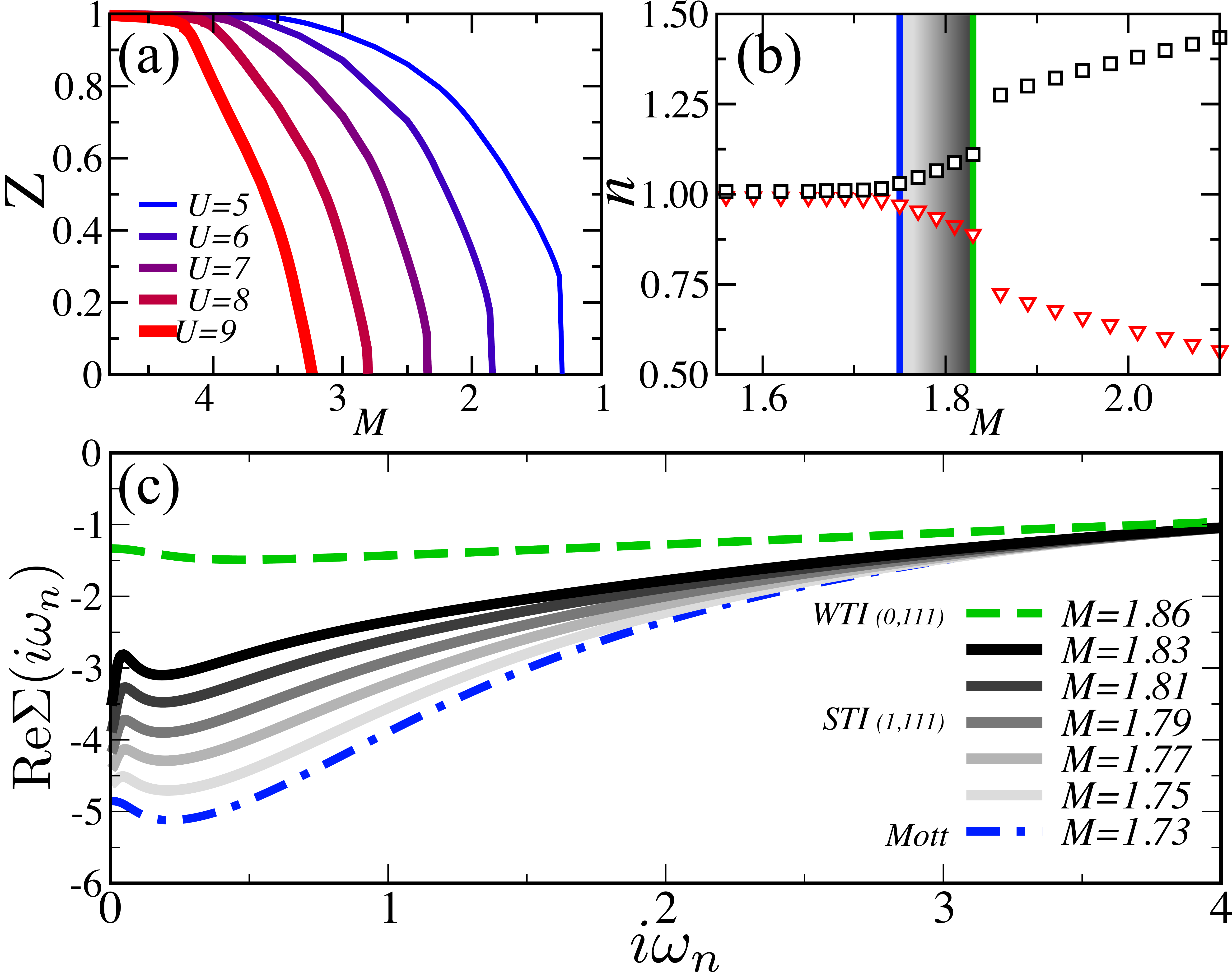}
  \caption{(Color Online)
    (a) Renormalization constant $Z$ as a function of $M$. Data are for increasing values of
    $U$ and for $J=U/4$. Formation of the Mott insulating state is
    signaled by the vanishing of $Z$. The transition is generically of
    the first-order. The discontinuity reduces approaching the triple
    point.
    (b) Orbital occupation
    across the WTI-Mott Insulator transition through the
    STI$_R$. First-order jump takes place at WTI-STI$_R$ transition.
    (c) Real part of the scalar Matsubara self-energy
    $\Re\Sigma(\iome)$ for increasing values of $M$. Data are for
    $U=6$ and $J=U/4$. The plot illustrates how large self-energy lead
    to the formation of the {\it anomalous} STI$_R$ phase.
  }
  \label{figMott}
\end{figure}

In correspondence to the divergence of the imaginary part of the
self-energy at the Mott point, the real part $\Re\hat{\bf
\Sigma}(\iome)$ becomes large and strongly frequency dependent (see \figu{figMott}(c)).
This behavior is necessary for the opening and the stabilization of
the Mott gap. The progressive localization of the electrons
obliterates the topological properties of the system, which are
related to the low-energy band structure.
In other words the large negative (for $J>0$) values of $\Re\hat{\bf \Sigma}(0)$
dominate the spectrum of the topological Hamiltonian
\equ{topological_hamiltonian}, giving rise to a trivial topological
invariant $\vec{\nu}=(0;000)$ for the Mott phase.

Interestingly, we find that for intermediate values of
both the interaction $U$ and mass $M$ the system admits an {\it anomalous}
topological phase in proximity of the Mott transition point.
In particular, our results show that the formation of the Mott
insulator can be preceded by a TQPT to a STI state with $\vec{\nu}=(1,111)$, \ie
STI$_R$, which coating the boundary line separates the WTI from the
Mott insulating region.
The existence of such tiny region near the Mott phase is reported in
the phase-diagram in \figu{figPD} (right panel).
The origin of such anomalous STI$_R$ state in the $M>0$ can
be understood from the behavior of the self-energy in proximity of the
Mott point. Near the Mott region  $\Re\hat{\bf \Sigma}(\iome)$ can attain
values which are large enough to compensate the initial $M$, but
such that the effective mass is pushed into
the range $M_\mathrm{eff}\in[-1,-3]$. For any such value of
$M_\mathrm{eff}$ the spectrum of the topological
Hamiltonian would correspond to a global topological invariant of
$\vec{\nu}=(1;111)$, \ie give rise to a STI$_R$ phase.

Unexpectedly, the anomalous STI$_R$ phase is separated from the WTI by
a first-order like transition, while it is continuously connected with
the Mott insulating region. This behavior is well visible in the
\figu{figMott}(b), where we report the orbital occupations as a function
of $M$ in the region near the anomalous STI$_R$ phase.
As the plot shows, for small $M$ the occupations are both equal and
identical to one, as expected in the Mott phase.
By increasing $M$ above the Mott transition value $M>M_\mathrm{MT}$ the occupations
continuously deviates from one, signaling the onset of the anomalous
STI$_R$ phase. This behavior is suddenly arrested by increasing $M$ above
a second critical value $M>M_\mathrm{R}$, at which a first-order
discontinuity in the occupations denotes the TQPT to the WTI phase.

\section{Conclusions}\label{secConclusions}
We studied the effects of strong electronic correlation on the
properties of a paradigmatic model for three-dimensional topological
insulators.
In particular, we considered a local density-density
{\emph{multi-orbital}} electronic repulsion, in presence of a Hund's coupling
$J$ taking into account the tendency of electrons to maximize the
total spin orientation while minimizing the orbital polarization.
We solved the interacting problem in a non-perturbative way using dynamical
mean-field theory. We determined the zero temperature phase-diagrams of the model as
a function of the interaction strength $U$ and the crystal-field, or
mass term, $M$, both for zero and finite values of the local Hund's'
coupling $J$.
We explained the specific form of the diagrams in both cases in terms
of the different behaviours of  the self-energy functions, accounting for
the interaction effects at the single-particle level. The
existence of a topologically trivial Mott insulating state for $J>0$
and large $U$ is shown.

We point out that, notably, the phase diagrams feature the presence of a critical point on the
topological quantum phase-transition line separating the trivial band
insulators from the strong topological phase.
In addition we unveiled the evolution of the topological transition
crossing such critical point. Our findings demonstrate that in the
weak-interaction regime the transition remains continuous while in the
strong-coupling regime, i.e. beyond the critical point, the
transition becomes discontinuous, i.e. a first-order transition.

The main consequence of the first-order character appearing beyond a
critical point in the orbital sector is the absence of gap
closing. This means that the inversion in the orbital character
responsible for the change in the topological invariants is not
accompanied by a continuous evolution of the spectral gap, if
many-body processes dominate.
This is a novel observation that clearly characterizes the transition
from the trivial band-insulator to the strong-TI phase. The successive
strong-to-weak topological transition is instead always of
second-order character, as in the non-interacting band-structure.
These results generalize and extend to the three-dimensional case our
findings for the topological transition to the quantum spin Hall state
in two-dimensions~\cite{Amaricci2015PRL}.

In addition, we find the existence of a $(1;111)$ phase, i.e. a strong TI phase
which appears before the eventual transition from the weak- to
the trivial Mott insulator. 
An analysis of the orbital occupations in
this narrow region of the phase diagram reveals that this phase is
separated from the WTI $(0;111)$ by a first-order transition and from
the Mott insulating state by a continuous one. 
Although this phase has the same topological indices $(1;111)$ as the
single-particle STI$_R$ phase at $U=0$ (i.e. $M\in[-1,-3])$, the two
can not be directly connected in our phase-diagram without crossing a
topological transition line. Indeed, the re-entrant $(1;111)$ phase emerges
here as the result of the strong many-body correlation encoded in the
dynamical self-energies. 

\subsection{Acknowledgments}
G.S. is grateful to Piet Brouwer for useful discussions.
M.C. and A.A. acknowledge financial support from the European Research
Council under FPO7 Starting Independent Research Grant n.240524 ``SUPER BAD".
B.T. and G.S. acknowledge financial support by the DFG (SPP 1666 and SFB 1170). J.C.B. acknowledges financial support from the ERC synergy grant UQUAM.

\bibliography{localbib}

\end{document}